\documentclass[twocolumn,showpacs,amsmath,amssymb,longbibliography,superscriptaddress,showkeys]{revtex4-1}
\usepackage{color,longtable}
\usepackage{graphicx}
\usepackage{amssymb,amsfonts,amsmath}
\usepackage{here}
\begin{document}
\title{Gel-like granular materials with high durability and high deformability}
\author{Honoka Fujio, Hikari Yokota, Marie Tani and Rei Kurita}

\affiliation{%
Department of Physics, Tokyo Metropolitan University, 1-1 Minamioosawa, Hachiouji-shi, Tokyo 192-0397, Japan
}%
\date{\today}

\begin{abstract}
Building materials such as concretes and mortar are formed by solidifying granular slabs. Such materials are often fractured by giant forces such as during earthquakes, leading to the collapse of structures and potentially casualties. One avenue of enquiry to prevent cracking would be to realize a material that can maintain a stable shape without being solidified. Here, we focus on sand grains coated with silicone oil, experimentally investigating the Young's modulus of a granular slab where ordinary grains and the coated grains are mixed in a mixing ratio $\alpha$. It is found that the Young's modulus increases rapidly at $\alpha \ge 0.6$. We use numerical simulation to show that this sudden increase in the Young's modulus is caused by a rigidity percolation transition. Furthermore, we are able to show that granular slabs containing coated sand have outstanding deformability without collapsing under large external stress. We believe this may lead to the development of granular materials that are rigid under usual pressures but deformable under more extreme conditions, such as during seismic activity. 
\end{abstract}

\maketitle
Granular systems are ubiquitously observed in daily life. Examples include sandboxes, dunes, flour, beads, and pharmaceutical powders. Despite individual particles obeying classical equations of motion, granular systems are known to exhibit complex macroscopic behavior. They can behave like a fluid, as in an hourglass, or like a solid, when they form arches or domes, clogging hoppers and pipes \cite{duran2012,brown2016,nagel1992,jaeger1992,guyon1994}. Understanding the behavior of granular systems is an important challenge for agriculture, construction, pharmaceuticals, and preparing for natural disasters such as landslides and avalanches. 

It is known that forces inside dry granular systems do not propagate uniformly but in a chain-like manner. This force chain is intricately linked to their macroscopic behavior \cite{drescher1972}. Although dry granular materials are well studied, it should be noted that granular materials found in nature and our daily lives often contain liquid. With the addition of even a small amount of liquid, the mechanical properties of granular materials are dramatically altered. The liquid bridges grains (liquid adhesion or capillary bridging), and the capillary force causes an effective attractive interaction between the particles~\cite{mitarai2006,samadani2000,scheel2008}. Therefore, it has greater rigidity than dry granular materials. Concretes, for example, are made by solidifying the liquid bridged state, and are found in many everyday applications, such as constructions and roads. Such materials do not have a relaxation time scale since they are solidified, and consequently cannot relax their internal structure. When gigantic perturbing forces such as earthquakes are applied, they can undergo brittle fracture, resulting in human casualties. Therefore, there is a need for novel granular materials that not only have high durability but also large deformability.

Sand coated with silicone oil (coated sand) is a product designed as a children's toy ~\cite{Kinetic}. The silicone oil is bonded directly to each grain and does not peel off. The coating allows the oil to persist stably around the sand. When the oil layers come into contact with each other, liquid adhesion occurs, causing an attractive interaction between coated particles~\cite{Tani2021}. 
Here we note that this attractive interaction remains for a long time, while the capillary bridging cannot retain due to evaporation or drainage.
Experiments where this coated sand was mixed with ordinary grains have been reported as a model system for wet granular materials~\cite{Tani2021}. In such systems, microscopy observations have confirmed that no stable liquid adhesion occurs between coated and uncoated sand, no matter how many times or how long they are mixed. It was reported that the packing fraction of random loose packing, the angle of repose, and the amount of residual material when the mixture is sieved change significantly when the mixing ratio of the coated sand to the total mass $\alpha$ is over 0.23~\cite{Tani2021}. This is close to the connectivity percolation value of $\alpha \sim 0.25$~\cite{stauffer1994}, suggesting that the change in physical properties may be due to percolation of clusters formed by coated sand.

Previous experiments focused on physical properties with a mixing ratio of $\alpha < 0.3$, where the granular slab cannot retain its shape without a container~\cite{Tani2021}. 
It does not become a block and cannot be used as a granular material.
In this study, we focus on the mechanical properties of randomly close packed (RCP) blocks of material as a possibility for actual material development. 
Here we investigate the $\alpha$ dependence of the Young's modulus and deformability of a granular block. 
We also seek to clarify the origins of large Young’s modulus and large deformability using numerical simulations.

For ordinary grains, we used Tohoku Silica Sand No. 5 sand (Tohoku Silica Sand Co., Ltd.). These grains have a density of 2.62 g/cm$^3$, a short diameter of 0.08-0.35 mm, and a long diameter of 0.23-0.55 mm. Kinetic Sand (Rangs Japan) was used for silicone oil coated sand. The density is 2.26 g/cm$^3$, the short diameter is 0.09-0.29 mm, and the long diameter is 0.16-0.41 mm. For each mass ratio $\alpha$, we measured a mass of each, spread them on the same metal vat and mixed them well by hand until uniform. Microscopic observation of several locations confirmed that the mixture was nearly uniform.

We prepared the random close packed (RCP) slabs in the following way. We gently put the sand into a cylindrical container with a diameter of 74.0 mm and height of 42.0 mm, or a rectangular container with a 67.8 mm $\times$ 67.8 mm base and a height of 42.5 mm. The rate at which the sand was poured was about 10 g/s. RCP states are generated by compacting mixtures down by external pressure ($\sim$ 5 $\times$ 10$^4$ Pa) using a plate whose area is almost the same as the inside of the container. We then poured more sand over the top until the container is full. We repeated this process around 25 times. The packing fraction of the granular slabs become roughly uniform. This state can be regarded as being RCP. 

\begin{figure}[h]
\begin{center}
\includegraphics[width=85mm]{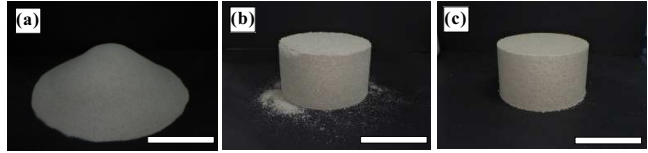}
\end{center}
\caption{Appearance of samples 60 minutes after they were removed from a cylindrical container, where (a) $\alpha$ = 0.0, (b) 0.3, and (c) 0.5. The scale bar corresponds to 50 mm. }
\label{image1}
\end{figure}

\begin{figure}[h]
\begin{center}
\includegraphics[width=85mm]{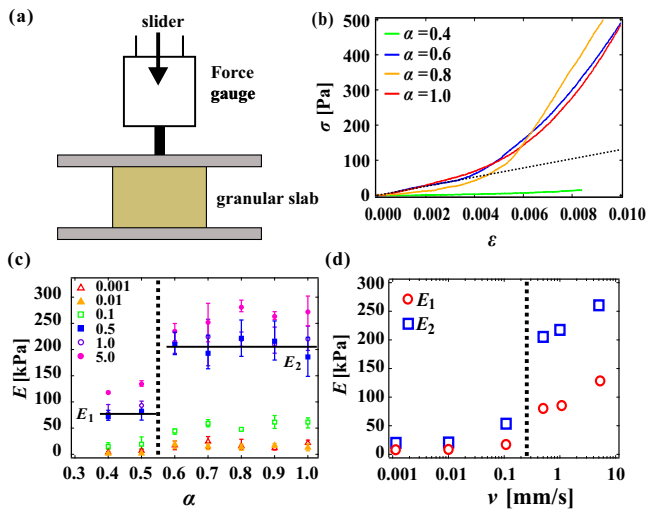}
\end{center}
\caption{(a) Schematic for the experimental setup, viewed from the side. (b) $\sigma$-$\varepsilon$ curve. When the strain is small, the normal stress increases linearly (dashed line) and the Young's modulus of the grain slab $E$ can be estimated as $E = \sigma/\varepsilon$ from the slope. (c) Young's modulus $E$ as a function of $\alpha$. The Young's modulus increases sharply at $\alpha \sim 0.6$ for all compression rates $v$. The mean values of the Young's modulus over 0.4 $\le \alpha \le$ 0.5 and 0.6 $\le \alpha \le$ 1 are labeled $E_1$ and $E_2$, respectively for each $v$. The solid lines represent $E_1$ and $E_2$ for $v$ = 0.5 mm/s. (d) The Young's modulus as a function of the compression rate $v$. $E_1$ and $E_2$ increase sharply by approximately 5 times at a transition compression rate of $v_t = 0.2$ mm/s.}
\label{Young}
\end{figure}

We flipped the container over and extracted the RCP granular slab from the container (see Fig.~\ref{image1}). Next, external pressure was applied by using a slider (LT-S300, Thorlabs) equipped with a large plate (Fig.~\ref{Young}(a)). The plate is designed so that it always stays horizontal. The slider can move with a constant velocity $v$ with a minimum $v$ of 0.001 mm/s. The force was measured using a force gauge (DST-50N, IMADA, Japan) during the slider movement. The resolution of the force is 0.001 N. We measured this 10 times using the same experimental conditions; the error bars correspond to the maximum and minimum found during the experimental trials. 

We performed overdamped numerical simulations to investigate the internal structures of granular slabs. To simplify, we assume grains to be spherical with a diameter $d_i$ for particle $i$. We set the size distribution to be Gaussian with a polydispersity of 20 \%. A two-body potential $U(r_{ij})$ between particles $i$ and $j$ is applied such that $U(r_{ij}) = \frac{k}{2} [r_{ij} - (d_i+d_j)/2]^2 $ if $(d_i+d_j)/2 > r_{ij}$, where $r_{ij}$ is the distance between particles $i$ and $j$ and $k$ = 1. Otherwise, $U(r_{ij}) = 0$. Firstly, we prepared a RCP state of non-coated grains using a method reported in Ref.~\cite{Xi2005, Clarke1987}. The particles are overpacked until the packing fraction is 0.645, and then shrunk until the total potential is below 10$^{-3}$. Next, we replace a certain number of non-coated sand grains with coated grains until the number ratio of coated grains becomes $\alpha$. Coated grains $i$ and $j$ interact via an attractive potential $U(r_{ij}) = \frac{k_{att}}{2} [r_{ij} - (d_i+d_j)/2]^2 $ when $(d_i+d_j)/2 < r_{ij} < r_c$, where $r_c$ is a threshold for the attractive interaction. We set $k_{att}$ = 0.1 and $r_c = 1.05 (d_i+d_j)/2$; 
we also confirm that our results are not sensitive to the threshold value.
We measured 3 different particle configurations and 5 different ways to replace uncoated grains with coated grains for each data point. 

Firstly, we examined the stability of cylindrical RCP samples under their own weight. Figure~\ref{image1} shows the appearance of the samples with $\alpha$ = 0.0, 0.3, and 0.5, 60 minutes after it was removed from a cylindrical container. When $\alpha \le$ 0.25, the sample collapses the moment it is removed from the container and forms a mound [Fig.~\ref{image1}(a)]. The angle of repose is 20$^\circ$ $\sim$ 25$^\circ$, which is almost the same as that of an ordinary sand pile~\cite{Tani2021,brown2016}. On the other hand, a slab with $0.3 \le \alpha \le 0.4$ only partially collapses as shown in Fig.~\ref{image1}(b), almost maintaining its shape. Finally, when the ratio of coated sand increases to $\alpha \ge 0.5$, the granular slab maintains its shape completely, as shown in Fig.~\ref{image1}(c). 
The slab keeps its shape perfectly and does not collapse after a long time. 

Next, we investigate response to large external forces. As shown in the experimental schematic in Fig.~\ref{Young}(a), samples were compressed from the top surface at a constant speed by a plate, and the normal stress $\sigma$ was measured. When $\alpha \le 0.3$, the sample collapsed immediately and $\sigma$ could not be measured. When $\alpha \ge 0.4$, they did not collapse under small deformations. Figure~\ref{Young}(b) shows the dependence of the normal stress $\sigma$ on strain $\varepsilon$ for $\alpha$ = 0.4, 0.6, 0.8 and 1.0 at a compression speed of $v$ = 0.01 mm/s. When the strain is small, the normal stress increases linearly; as the strain is increased further, the stress increases rapidly. However, we see that the $\sigma$-$\varepsilon$ curve is qualitatively different for $\alpha$ = 0.4 and $\alpha \ge 0.6$. Here, we measured the slope of the linear region under small strains and estimated the Young's modulus of the grain slab, $E = \sigma/\varepsilon$, from the slope. Figure~\ref{Young}(c) shows the relationship between $\alpha$ and $E$. The Young's modulus changes rapidly for $0.5 < \alpha < 0.6$ and then remains almost constant for $\alpha \ge 0.6$.

The same experiment was conducted at different compression speeds. As shown in Fig.~\ref{Young}(c), the Young's modulus increases sharply at $\alpha \sim 0.6$ for all $v$. The means of the Young's modulus for 0.4 $\le \alpha \le$ 0.5 and at 0.6 $\le \alpha \le$ 1 are calculated as shown in Fig.~\ref{Young}(c) and labeled $E_1$ and $E_2$, respectively for each $v$. $E_1$ and $E_2$ increased sharply by several times at a transition compression rate $v_t \sim 0.2$ mm/s as shown in Fig.~\ref{Young}(d). This suggests the existence of a relaxation time scale for the internal structure. The characteristic time may be estimated by $\tau \sim \langle d \rangle/v_t ~ \sim$ 1 s, where $\langle d \rangle$ is the mean size of the grains. A possible explanation for this time scale can be the lifetime of the attractive interaction.
The time scale over which this liquid bridge disconnects is about 1 s for our coated grains. 
Another possibility is structural relaxation, where particles in contact rearrange in the same way as $T1$ events in foams~\cite{Yanagisawa2021}. We are also working on a detailed study of the origin of this time scale and will report these results in the near future.

\begin{figure}
\begin{center}
\includegraphics[width=85mm]{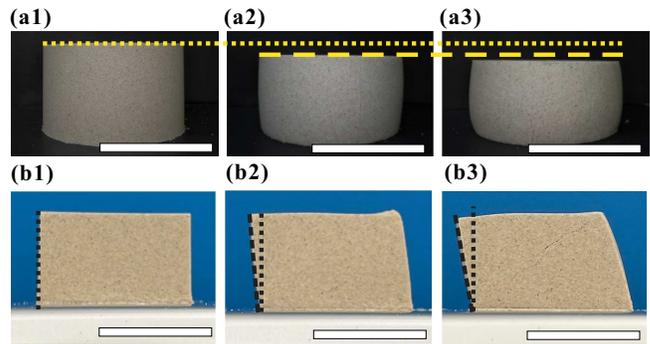}
\end{center}
\caption{Snapshots of a granular slab with $\alpha = 0.6$ under different strains, $\varepsilon = $ (a1) 0.0, (a2) 0.10, and (a3) 0.15. Dotted and dashed lines show the positions of the top surface at $\varepsilon$ = 0.0, and 0.1, respectively.  Snapshots of a rectangular granular slab with $\alpha = 0.6$ under different shear strains $\varepsilon = $ (b1) 0.0, (b2) 0.10, and (b3) 0.15. Dotted vertical lines shows the perpendicular to the substrate and dash lines show the tilted edge of the granular slab. The granular slab can be deformed without cracking even though $\varepsilon = 0.15$ and the granular slab is stable at least for an hour.
The scale bar corresponds to 50 mm. }
\label{compress}
\end{figure}

Here, we note the importance of the granular slab having a structural relaxation in the context of construction materials. Normal concretes undergo brittle fracture when they are deeply compressed or a crack emerges inside. However, granular slabs with $\alpha \ge 0.6$ can be deformed without cracking even when $\varepsilon = 0.15$, as shown in Fig.~\ref{compress}(a1)-(a3); it is found that the width at the center of the granular slab expanded more than at the base. This means that the structure may have changed inside the slab, something which cannot occur when the slab is solidified.  
We also check the deformability for the shear deformation. 
We sandwiched the granular slab by two acrylic plates and then we sheared the granular slab by tilting the acrylic plates. 
The granular slabs with $\alpha \ge 0.6$ can be deformed even when $\varepsilon = 0.15$, as shown in Fig.~\ref{compress}(b1)-(b3).
The granular slabs with $\varepsilon = 0.15$ is stable at least for an hour.

\begin{figure}[h]
\begin{center}
\includegraphics[width=85mm]{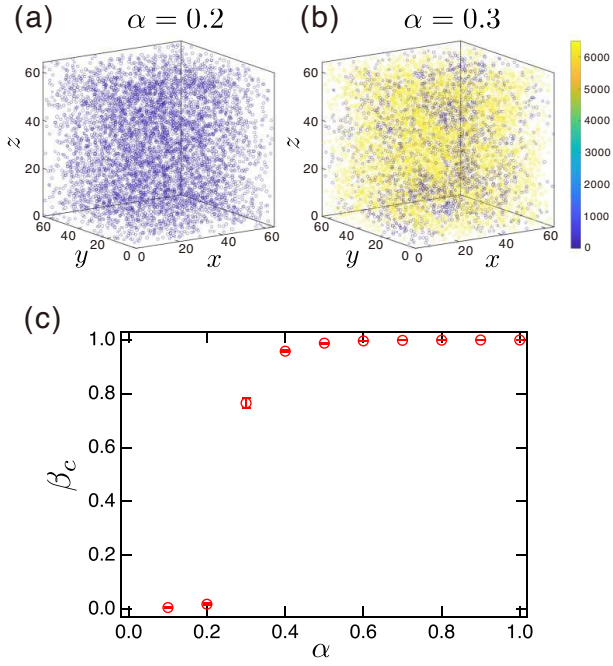}
\end{center}
\caption{Spatial distribution of clusters composed of connected coated sand grains when (a) $\alpha$ = 0.2 and (b) 0.3. The colors correspond to the size of each cluster. It is found that many small clusters are scattered over space when $\alpha$ = 0.2, while one large cluster percolates the system at $\alpha$ = 0.3. (c) $\beta_c$ as a function of $\alpha$. The maximum cluster size increases significantly at $\alpha = 0.3$. The error bars correspond to the standard deviations.}
\label{connectivity}
\end{figure}

Next, to clarify the origin of the change of the Young’s modulus, we compute the inner structures using a numerical simulation of packings spheres which model our mixture; details given in the Methods section. Firstly, we focus on connections between coated sand grains. Figure~\ref{connectivity} shows the spatial distribution of clusters composed of coated sand grains at (a) $\alpha$ = 0.2 and (b) 0.3. The colors represent the size of each cluster. It is found that many small clusters are scattered when $\alpha$ = 0.2, while there is one large cluster when $\alpha$ = 0.3. We measured the particle number in each connected cluster $N_c$ and defined $N_c^{max}$ as the number of particles in the largest cluster. Figure~\ref{connectivity}(c) shows the ratio of $N_c^{max}$ to the total number of coated sand particles $\beta_c$, that is, $\beta_c = N_c^{max}/N \alpha$. It is found that $\beta_c$ increases significantly at $\alpha = 0.3$. As $\alpha$ increases further, $\beta_c$ approaches 1; this means that almost all coated sand grains are included in one large cluster. We also confirm that the large cluster percolates the system. Furthermore, the contact number $Z$ in the maximum cluster is $Z$ = 2.77 at $\alpha = 0.3$. It has been reported previously that percolated structures are rigid when $Z \ge 2.4$~\cite{Kuo1995,Javier2021}. Thus, the connectivity percolation transition corresponds to the shape-retaining property of the granular slab shown in Fig.~\ref{image1}. This is also consistent with the correspondence of the connectivity percolation with changes in the packing fraction of random loose packing, behavior under sieving, and the angle of repose as reported in Ref.~\cite{Tani2021}.

\begin{figure}[h]
\begin{center}
\includegraphics[width=85mm]{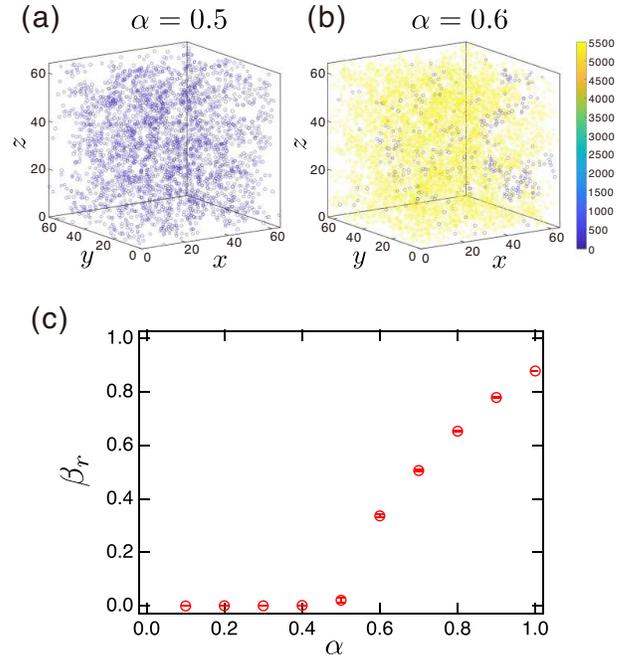}
\end{center}
\caption{Spatial distribution of rigid clusters where the contact number is over 6 when (a) $\alpha$ = 0.5, (b) $\alpha$ = 0.6. The color represents the size of each rigid cluster. It is found that many small rigid clusters are scatter over space at $\alpha$ = 0.5, while one large rigid cluster percolates the system at $\alpha$ = 0.6. (c) $\beta_r$ as a function of $\alpha$. Note that $\beta_r \sim$ 0.3 at $\alpha = 0.6$; this is larger than the value required for percolation. The error bars correspond to the standard deviations.}
\label{rigidity}
\end{figure}

The connectivity percolation should increase the stiffness of the granular slab, but it is also known that connectivity is not enough to realize mechanical rigidity~\cite{maxwell1864, guyon1990,aharonov1999, Lois2008, Yanagishima2021}. For example, when a particle makes contact with two particles, a linear cluster can be formed. Such a cluster is weak against large stresses and can be easily deformed. Meanwhile, Maxwell's condition stipulates that connections become rigid when the total contact number $z$ reaches $6n$-12, where $n$ is the particle number~\cite{maxwell1864}. 
Therefore, the contact number for each particles is almost 6 since $n$ is enough large. 
We defined a rigid connection if the contact number is over 6. 
We apply this to the structures generated in our simulation and investigate the spatial distribution of rigid connections. Figure~\ref{rigidity} shows the distribution of rigid clusters at (a) $\alpha$ = 0.5 and (b) 0.6. Now, colors correspond to the size of rigid clusters. It is found that small rigid clusters are scattered over space at $\alpha$ = 0.5, even though the clusters as judged by connections are percolated. This means that the connective clusters are deformable when a large external force is applied. 
Meanwhile, when $\alpha$ = 0.6, it is clear that there is one large rigid cluster. We measured the number of particles in each rigid cluster $N_r$ and define $N_r^{max}$ as the particle number in the largest cluster. Figure~\ref{rigidity}(c) shows the ratio of $N_r^{max}$ to the total number of coated particles $\beta_r$, that is, $\beta_r = N_r^{max}/N \alpha$. It is found that $\beta_r$ becomes non-zero at $\alpha = 0.6$; the value $\beta_r$ = 0.3 is larger than that required for percolation; we also directly confirm that the rigid cluster percolates the system by checking the coordinates of the particles. This rigidity percolation is the reason why the Young’s modulus suddenly increases for $\alpha \ge 0.6$. 

Here, the mechanical properties of this mixed granular slab strongly depend on the connection between coated sand particles via an attractive interaction. This has a close analogy with gels, where polymer networks may percolate \cite{okumura2001, sakai2008}. Meanwhile, we note that granular systems are generally related to jammed system. Thus, this mixed granular system may have the physical properties of both gels and granular systems, such as large rigidity and large deformability without brittle fracture. This may mean that new functional granular materials may be produced by simply changing the coating of a certain proportion of grains. 

We investigated a unique granular slab composed of a mixture of sand grains coated and not coated with silicone oil. We measured the mechanical properties of RCP granular slabs as a function of the mass ratio of coated sand grains $\alpha$. It is found that granular slabs can retain their own shape when $\alpha \ge 0.3$. We also found that the Young's modulus suddenly increases at $\alpha$ = 0.6. From the results of numerical simulations, the coated grains not only percolate via connectivity at $\alpha \ge 0.3$, but also undergo rigidity percolation when $\alpha \ge 0.6$. The former percolation is related with shape retention, while the latter is correlated with the increased Young's modulus. Furthermore, the Young's modulus depends on a compression velocity, suggesting the granular slab has a characteristic relaxation time scale. Thus, we can expect viscoelastic behavior in our granular slabs, paving the way for new functional granular materials simply by changing the coating of a proportion of grains.


M. T. was supported by JSPS KAKENHI (20K14431) and R. K. was supported by JSPS KAKENHI (17H02945 and 20H01874). 
Correspondence and requests for materials should be addressed to R.~K. (kurita@tmu.ac.jp).


 
%

\clearpage
\end{document}